# Order-disorder transitions in a polar vortex lattice

Linming Zhou[1, #], Cheng Dai [2, #], Peter Meisenheimer[3], Sujit Das[3], Yongjun Wu[1, 4, *], Fernando Gómez-Ortiz[5], Pablo García-Fernández[5], Yuhui Huang[1], Javier Junquera[5], Long-Qing Chen[2, *], Ramamoorthy Ramesh[3, *] , Zijian Hong[1, 4, *]

[1] Lab of Dielectric Materials, School of Materials Science and Engineering, Zhejiang University, Hangzhou, Zhejiang 310027, China

[2] Department of Materials Science and Engineering, The Pennsylvania State University, University Park, PA 16802, USA

[3] Department of Materials Science and Engineering, University of California, Berkeley, California 94720, USA.

[4] Cyrus Tang Center for Sensor Materials and Applications, State Key Laboratory of Silicon Materials, Zhejiang University, Hangzhou, Zhejiang 310027, China

[5] Departamento de Ciencias de la Tierra y Física de la Materia Condensada, Universidad de Cantabria, Cantabria Campus Internacional, Avenida de los Castros s/n, E-39005 Santander, Spain

**Order-disorder transitions are widely explored in various vortex structures in condensed matter physics, i.e., in the type-II superconductors and Bose-Einstein condensates. In this study, we investigated the ordering of the polar vortex phase in $[Pb(Zr_{0.4}Ti_{0.6})O_3]_n/(SrTiO_3)_n$ (PZT/STO) superlattices through phase-field simulations. With a large tensile substrate strain, an antiorder vortex state (where the rotation direction of the vortex arrays in the neighboring ferroelectric layers are flipped) is discovered for short-period PZT/STO superlattice. The driving**

**force is the induced in-plane polarization in the STO layers due to the large tensile epitaxial strain. Increasing the periodicity leads to the antiorder to disorder transition, resulting from the high energy of the head-to-head/tail-to-tail domain structure in the STO layer. On the other hand, when the periodicity is kept constant in short-period superlattices, the order-disorder-antiorder transition can be engineered by mediating the substrate strain, due to the competition between the induction of out-of-plane (due to interfacial depolarization effect), and in-plane (due to strain) polarization in the STO layer. The 3-dimensional ordering of such polar vortices is still a topic of significant current interest and we envision this study will spur further interest towards the understanding of order-disorder transitions in ferroelectric topological structures.**

[#] Equal contributions

[*] Corresponding authors:

Y. W. (yongjunwu@zju.edu.cn); L. C. (lqc3@psu.edu) ;
R.R. (rramesh@berkeley.edu); Z. H. (hongzijian100@zju.edu.cn)

The order-disorder phase transition is a fundamental physics and materials phenomenon in nature. In condensed matter physics, it has been extensively investigated in a variety of topological systems, including vortices in high-temperature type-II superconductors [1-3], and rotating Bose-Einstein condensates [4-6]. For instance, it is widely recognized that vortex pinning and thermal fluctuation could lead to the order-disorder transition in the vortex phase of a superconductor, leading to the formation of a *vortex glass* state [7], which in turn strongly affects the critical current and threshold magnetic field of the superconducting phase.

In ferroelectric systems, various topological phases such as vortices [8-15], merons [16], flux-closure domains [17-19], spirals [20] and skyrmions [21-24] have been observed in the low-dimensional ferroelectric heterostructures in recent years. These topological structures usually arise from the complex interplay of charge and lattice degrees of freedom, as well as the depolarization and surface effects within the reduced dimensions. One particular example is the observation of a polar vortex lattice with the continuous rotation of polarization vector surrounding a singularity-like vortex core in the $(PbTiO_3)_n/(SrTiO_3)_n$ ($n$=10, 16) (PTO/STO) superlattice systems [11], akin to the Abrikosov vortex lattice [25, 26] observed in the type-II superconductors and ferromagnetic vortices. Previously, Yadav et al. [11] demonstrated that for the polar vortex in the PTO/STO superlattices, while the horizontal long-range ordering is robust, it has certain degree of vertical disorder among different PTO layers. This could potentially be a rich area of research that is largely unexplored [27]. One natural question would be: is it possible to design the vertical ordering of the polar vortex in the ferroelectric/dielectric superlattice systems?

To address this question, we first define three possible ordering prototypes and establish a parameter (namely the "degree of order") to quantify the vertical ordering. **Fig. 1(a)-(c)** shows the schematics of three scenarios for the vertical alignment of the vortex arrays. The ideal, vertically ordered vortex structure is illustrated in **Fig. 1(a)**, where the vortices in the adjacent ferroelectric layers show the same rotation directions throughout the thickness of the superlattice film. In this scenario, each ferroelectric layer is identical within the superlattice film. In another case, as shown in **Fig. 1(b)**, the alignment of the vortex along the out-of-plane direction is random, i.e., there is barely any long-range vertical ordering or strong correlation of the vortices in the adjacent ferroelectric layers, giving rise to a disordered, *vortex-glass-like* state. **Fig. 1(c)** shows the vertical "antiorder" state, where the rotation direction of the vortex arrays in the neighboring ferroelectric layers are flipped, resulting in alternating clockwise/counterclockwise rotations between the adjacent layers. To further distinguish the three possible states, we define a "degree of order" parameter to quantify the ordering of the polar vortices, which is calculated by the mean of the cosine of the angle (θ) between the polarization vectors at the same position within the ferroelectric in neighboring layers (**Fig. 1d**). Under this definition, the degree of order for an ideal ordered state is 1, while for a completely random disordered structure it has a value of 0 and the antiorder state has a degree of order of -1. Thus, this parameter could help to identify the status of the vertical ordering of the vortex lattice. A similar parameter is also adopted by Takenaka *et al.* to quantify the degree of correlation in relaxor ferroelectrics [28].

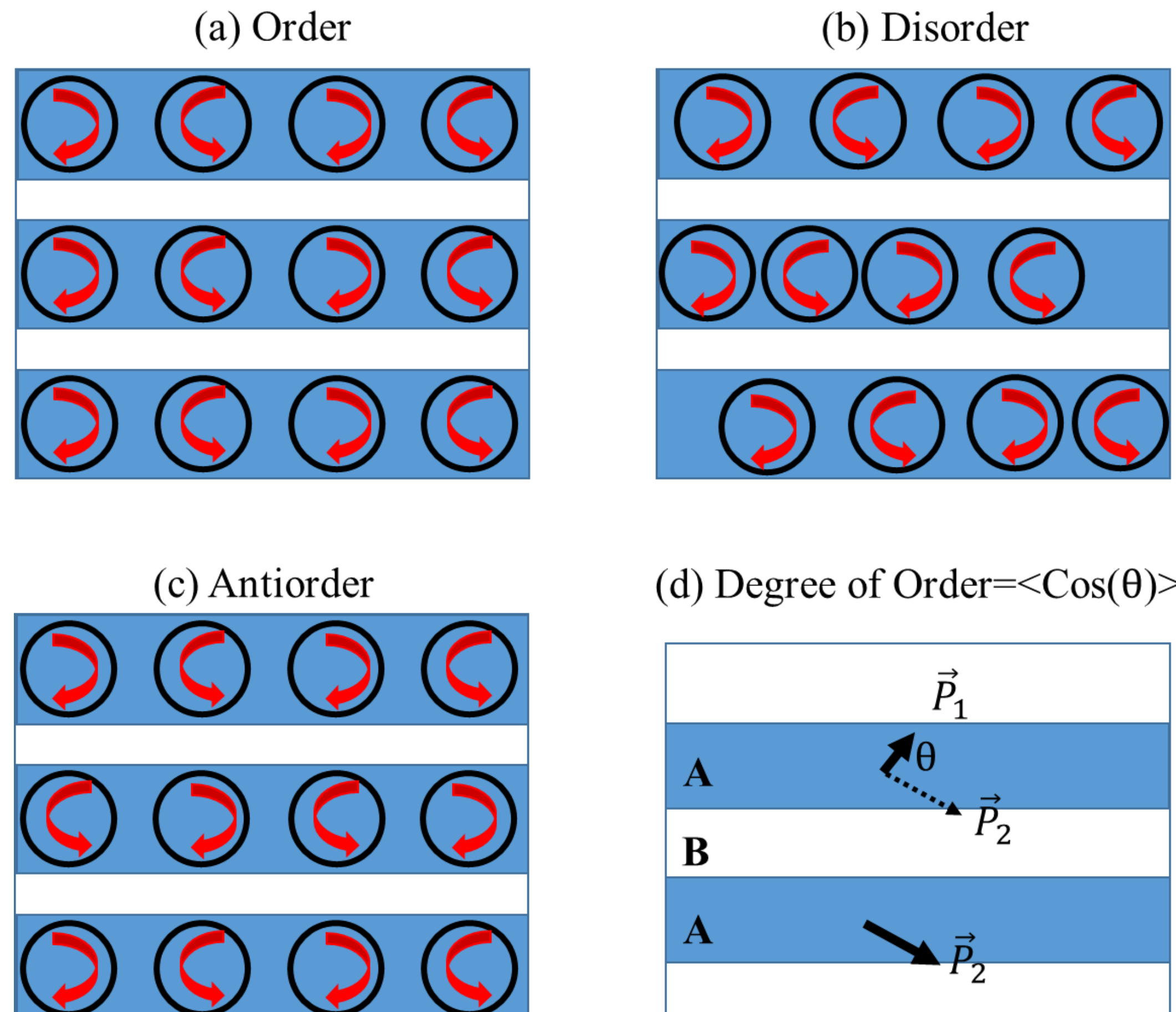

**Fig. 1| Schematics of different types of vertical order for the vortex system.** (a) Order, (b) disorder, (c) antiorder, and (d) definition of degree of order. The blue and white regions indicate the ferroelectric and dielectric layers, respectively. A black circle represents one polar vortex, where its rotation direction is indicated by the red arrow. Vertical order and antiorder indicate that vortices in the adjacent layers have the same or opposite rotation directions, while in disorder state there's no long-range correlation between the two layers. The degree of order is defined as the mean value of the cosine of the correlation angle. The degree of order value for the order, disorder and antiorder states is 1, 0 and -1, respectively.

Within the perovskite oxide family, to date, such polar textures have been studied almost exclusively in PTO/STO based superlattices [11-19, 22-24]. In this study, the $Pb(Zr_{0.4},Ti_{0.6})O_3/SrTiO_3$ (PZT/STO) superlattice is chosen as the model system. The role of isovalent doping in the Ti site (for example with $Zr^{4+}$) is well established through the PZT phase diagram [29-31]. For the Ti-rich part of this phase diagram, the structure remains tetragonal, but exhibits a systematic change in the lattice dimensions with Zr-content; particularly, the *a*-axis of the tetragonal unit cell progressively increases with Zr-substitution, while the *c*-axis decreases, resulting in a decrease in the polarization and increase in dielectric/piezoelectric responses. Thus, within the epitaxial superlattice framework, the PZT system provides a pathway to impose an additional source of strain, namely through the chemical pressure by Zr doping. As compared to the PTO/STO superlattice, PZT and STO have a larger lattice mismatch, which suggests the possibility for a wider strain-tuning window. Furthermore, PZT has a lower Curie temperature and smaller spontaneous polarization as compared to PTO, which suggests a more "flexible" polar state that could possibly be engineered to exhibit various topologies as well as degrees of ordering within them.

In this work, we investigate via phase-field simulations the phase-diagram of the polar structures of PZT/STO superlattices as a function of (i) its periodicity when they are assumed to be grown on a $(110)_o$-$SmScO_3$ (SSO) substrate; and (ii) when the epitaxial strain is changed from compressive to tensile, keeping the periodicity constant.

The anisotropic in-plane lattice constants of $(110)_o$-$SmScO_3$ (SSO) substrate are set as 3.991 Å and 3.983 Å [32], which causes a compressive strain in the PZT layer and tensile strain in the STO layer (the pseudocubic lattice constants are set as 4.040 Å [33] and 3.905 Å [15] for PZT and STO, respectively). Under this strain condition, an in-plane polarization is induced for the thick STO films, forming single *a* domain with polarization pointing toward the in-plane direction (**Fig. S1a**), in good agreement with previous reports [34]. On the other hand, for thick PZT thin films, *c* domains are observed with 180° $c^+/c^-$ domain walls (**Fig. S1b**). The induced polarization for the STO films is ~0.2 C/m$^2$, while PZT thin films show much larger polarization ~0.6 C/m$^2$ (**Fig. S1c**).

Our simulations suggest that, on the one hand, for the $(PZT)_n/(STO)_n$ on a $(110)_o$-$SmScO_3$ (SSO) substrate, a vertical antiorder vortex structure can be engineered with small periodicity (e.g., *n*=6), which is triggered by the in-plane polarization in the STO layers induced by the tensile substrate strain. The antiorder to disorder transition is driven by the increase of the periodicity of PZT/STO superlattice on a $(110)_o$-$SmScO_3$ (SSO) substrate, owing to the increase of *c*-oriented regions in the PZT layers, as well as to the requirement of reducing the high electrostatic energy associated with the head-to-head/tail-to-tail domain walls in the STO layers. On the other hand, when SSO is replaced by another substrate that imposes a compressive strain, an ordered vortex lattice state can be formed through the out-of-plane polarization of the STO layer. A strain phase diagram further indicates a strain-mediated order/disorder/antiorder transitions by increasing the in-plane lattice constant of the substrate.

As a mesoscale simulation tool, phase-field simulation has been widely adopted to study the polar topological structures and topological phase transitions in the ferroelectric heterostructures, and have successfully predicted several topological structures including polar vortices [11, 12], polar merons [16], polar spirals [20], and polar skyrmions [22-24]. The spontaneous polarization vector ($\vec{P}_i$, $i$=1, 3) is chosen as the order parameter, governed by the time-dependent Ginzburg-Landau (TDGL) equation [35-37]:

$$\frac{\partial \vec{P}_i}{\partial t} = -L\frac{\delta F}{\delta \vec{P}_i}\ (i = 1, 3) \qquad \text{(Eq. 1),}$$

where $t$ is the evolution time, $L$ is the kinetic coefficient related to the mobility of the ferroelectric domain wall, $F$ is the total free energy of the system which has the contribution from the Landau chemical, elastic, electrostatic, and polarization gradient energies, namely:

$$F = \int (f_{Landau} + f_{elastic} + f_{electric} + f_{gradient})dV \qquad \text{(Eq. 2).}$$

Detailed descriptions of the individual energies can be found in previous reports. [15, 35-38] The materials parameters (including Landau potentials, elastic constants, and background dielectric constant) are adopted from previous literature [15, 29-31, 39-41]. The simulation temperature for all the studies is set as 300 K. A three-dimensional mesh of 200×200×250 grids is established, with each grid point representing 0.4 nm. A periodic boundary condition is applied along the in-plane dimensions while the superposition method is applied along the out-of-plane dimension [42]. Along the thickness direction, the whole film is composed of 30 grid points of the substrate, 192 grid points of the (PZT/STO) film, and 28 grid points of air, respectively. The mechanical boundary condition of a thin film is applied, where the displacement at the bottom of the system is fixed to zero, while the out-of-plane stress on the top of the film is fixed to zero [36]. To account for the elastic inhomogeneity of the superlattice system, an iteration-perturbation method is used [43]. A short-circuit electric boundary condition is used, i.e., the electric potential at the top and bottom of the film is set to zero [38].

*Influence of the periodicity for a constant epitaxial strain.* The polar structure for the $(PZT)_6/(STO)_6$ superlattice on an SSO substrate is shown in **Fig. 2**. Long-range ordered polar vortices with a size of 2 nm form in the PZT layers (**Fig. 2a-b**), similar to the polar vortex lattice observed in the PTO/STO superlattices, which is due to the delicated balance of elastic, electric, Landau and gradient energies, as has been discussed in a previous report [15]. The planar view image (**Fig. 2c**) shows the formation of long vortex stripes, with a dislocation-like pattern, consistent with the vortex lattice observed in the PTO/STO system. Notably, it is revealed that the simulated structure exhibits antiorder, where the vortices in the two neighboring PZT layers have opposite rotation directions. This is largely due to the in-plane polarization in the STO layers induced by the tensile substrate strain, which could act as a "gear" where the in-plane polarization directions at the bottom of the vortices in the upper PZT layers share the same polarization directions with the top of the vortices in the lower PZT layers. This geometric confinement triggers the formation of a vertically aligned antiorder vortex structure. Since the in-plane polarization points in opposite directions at consecutive vortices, head-to-head and tail-to-tail domains appear within the STO layer, which costs energy. In order to minimize the electrostatic energy, the magnitude of the polarization gradually decreases and its direction bends to match the *c*-domains in the PZT layer. As a result, an antivortex structure (winding number -1) is formed in the STO layer, whose core lies right at the center of the square vortex lattice (**Fig. 2b**). Similar antivortex state has been simulated or observed in ferromagnetic nanostructures (e.g., in Permalloy dot [44] and $Fe_{19}Ni_{81}$ thin film [45]), and has also been observed in the short period PTO/STO superlattice [14, 46].

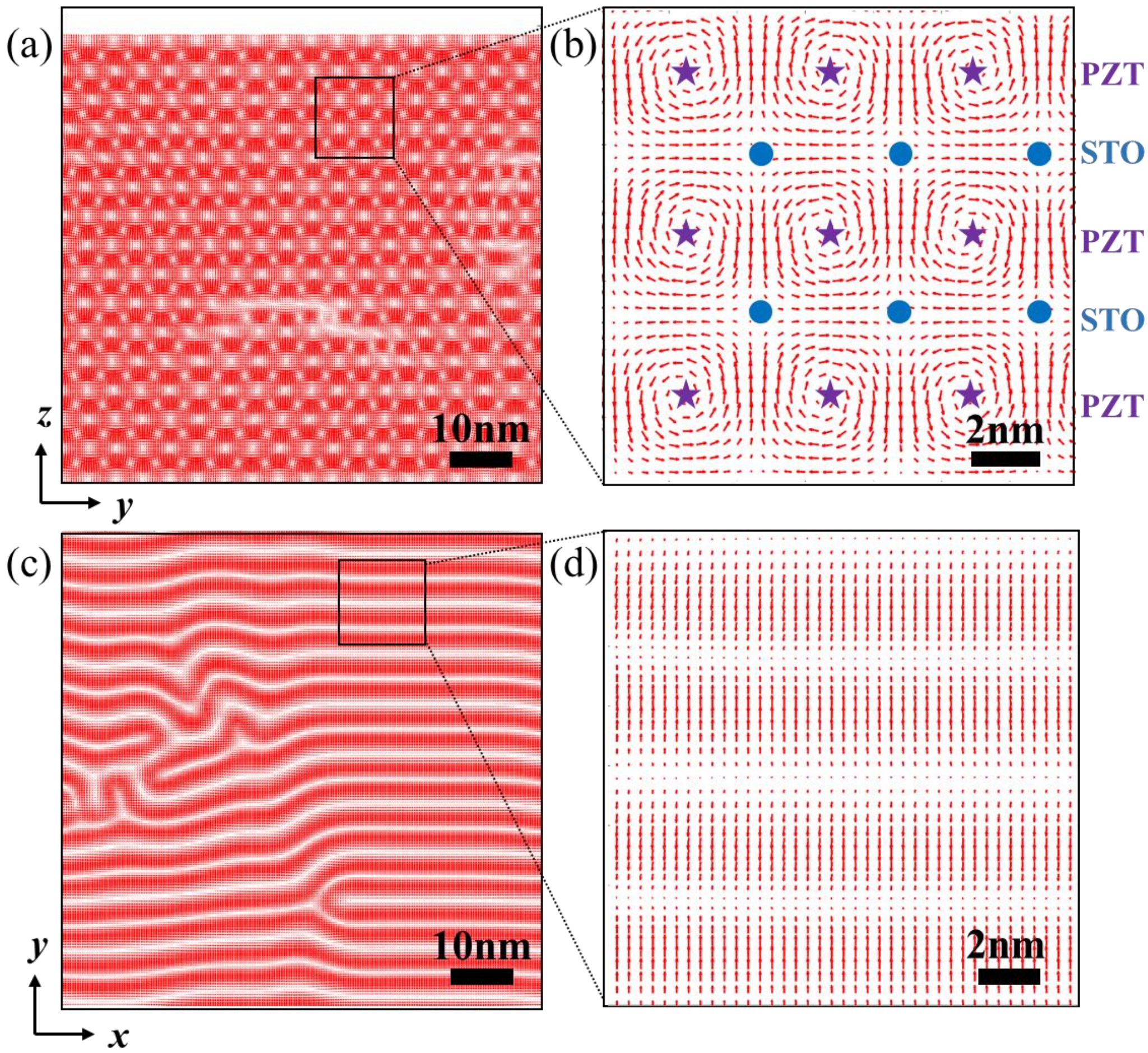

**Fig. 2| Polar structure for the $(PZT)_6/(STO)_6$ on a SSO substrate.** (a) Cross-section view of the polar vector, showing the alignment of the polar vortices. (b) zoom-in view. The core of the antivortex within the STO layer is marked with a blue dot, while the vortex core is marked with purple star. (c) Planar view on the top PZT layer, showing the formation of vortex stripes and dislocation patterns. (d) Magnified view of the planar view of the vortex stripes.

To further address the role of the STO layers on the rotation patterns and vortex vertical ordering for the PZT layers, the polar structure for $(PZT)_6/(STO)_n$ superlattice on SSO is further investigated (**Fig. S2**- **S5**). It can be observed that with the increasing of *n*, the energy of the head-to-head and tail-to-tail domains increases up to a point where they cannot be stable anymore. At this point (*n*~20), a phase transition is observed to form single in-plane *a* domains inside the STO layers (**Fig. S5**). Interestingly, this monodomain phase in the STO precludes the formation of vortices in the PZT layer, as clearly shown by the small density of these singularities in **Fig. S5**. For an intermediate thickness of STO, the vortex line in the PZT layer rotates by 90°.

We then explore the room temperature phase diagram for $(PZT)_n/(STO)_n$ superlattice epitaxially grown on a $(110)_o$-$SmScO_3$ (SSO) substrate as a function of the periodicity, (*n*=5-30) (**Fig. 3**). An antiorder to disorder transition is observed, with the degree of order switching from -1 to 0, with increasing periodicity. The driving force behind this evolution is twofold. First, the larger the periodicity, the larger the electrostatic energy that would be stored in the head-to-head and tail-to-tail domain walls within the STO layer. One way to release this electrostatic energy is to tilt these domain walls, in such a way that the polarization charge is reduced. This is observed in **Fig. 3(b)** for *n*=16, and translates into the fact that the in-plane components of the polarization within the STO layer connect vortices that are not aligned along the *z*-direction. Second, the larger the periodicity, the larger the out-of-plane polarization within the PZT layer. Indeed, for short-period superlattices (*n*=6), $P_z$ in the PZT layer is greatly suppressed, of the same magnitude as the strain-induced in-plane polarization in the STO layer (~ 0.2 $C/m^2$). For higher periodicities (*n*=16; **Fig. 3b**), the shape of the vortices has changed to a more rectangular domain wall type, while $P_z$ within the PZT layer increases up to ~ 0.4 $C/m^2$, with a concomitant increase of the coupling with the out-of-plane polarization in the STO layer (**Fig. S6b**). Therefore, to minimize the gradient of the polarization along the *z*-direction and the formation of polarization charges, there is a tendency to align the up/down domains in the PZT and the STO layer, favouring the appearance of a more ordered phase, and reducing the average value of the cosine shown in **Fig. 1** from its initial value of -1. For very large periodicities, the out-of-plane polarization within the STO layers cannot be maintained, and the PZT layers are totally decoupled.

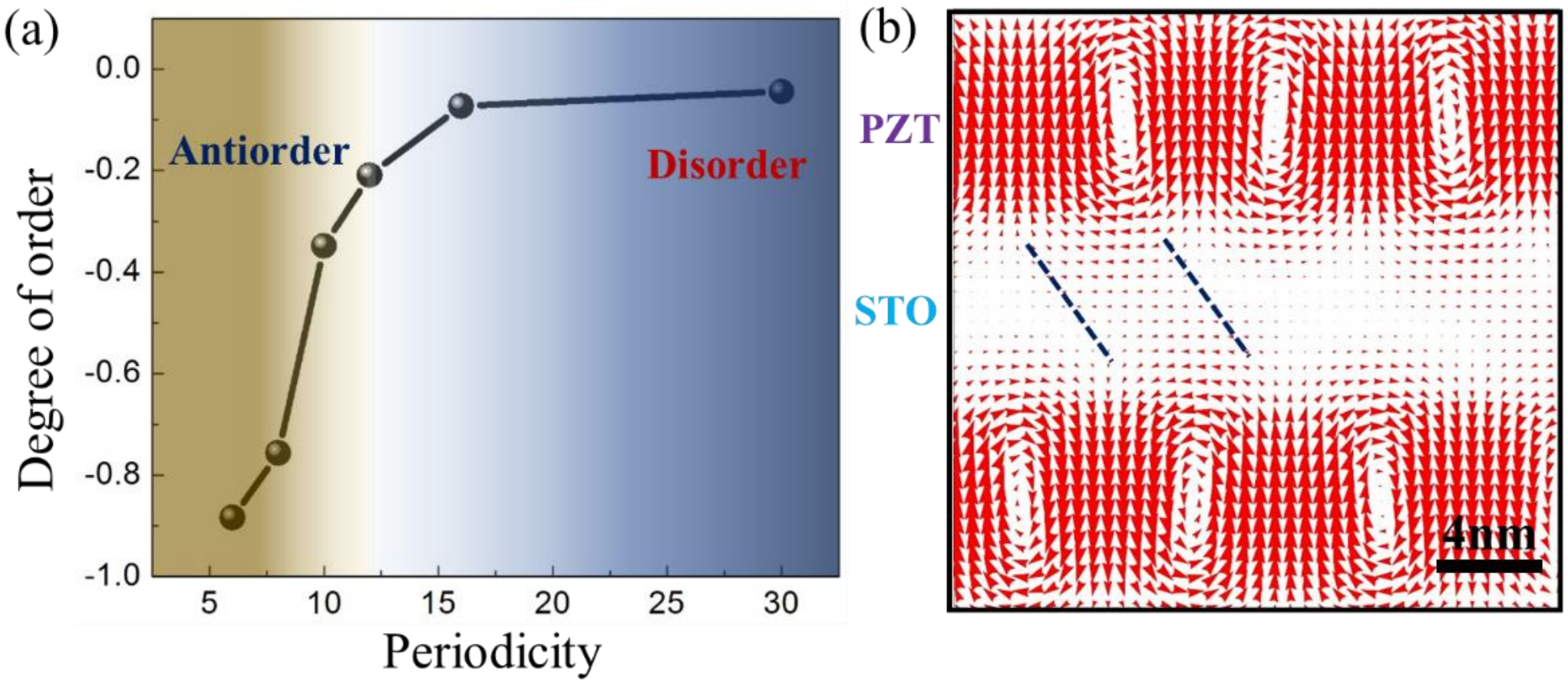

**Fig. 3| Periodicity phase diagram for $(PZT)_n/(STO)_n$ ($n$=5-30) superlattice on a SSO substrate**. (a) Degree of order as a function of periodicity, demonstrating the thickness-driven antiorder/disorder transition, dark yellow region indicates the antiorder structure, while dark blue shows the formation of disorder structure, where the light color demonstrates the transitional state in between a fully antiorder and disorder state. (b) Disorder structure for large periodicity $n$=16. The blue dashed lines marks the tilting of the head-to-head anf tail-to-tail domain walls.

*Influence of the epitaxial strain for a constant periodicity.* Since the polarization in the STO layer plays an important role in determining the degree of order, it can be expected that the epitaxial strain could be a key factor to tune it. The strain vs. degree of order diagram with fixed small periodicity ($n$=6) is plotted in **Fig. 4**. Three distinct regions can be distinguished. When the substrate lattice constant is small (<3.960 Å), an order structure can be observed with a degree of order close to 1. With increasing of the substrate lattice constant (from 3.960 Å to 3.975 Å), the compressive strain in the PZT layers decreases while the tensile strain in the STO layers increases, giving rise to the disordered state with a degree of order close to 0. Further increasing of the substrate lattice constant leads to the formation of an antiorder state, where the degree of order is close to -1.

When the PZT layer is under a large compressive strain, the out-of-plane polarization determines the physical behaviour of the superlattice. The polarization pattern in **Fig. 4(b)** shows the order structure where the vortices in the neighboring PZT layers rotate in the same direction. A finite out-of-plane polarization is observed in the STO layers, avoiding the discontinuity in the polarization and the appearance of electrostatically expensive bound charges. It shares the same out-of-plane polarization direction with the upper and lower PZT layers, thus tuning the order alignments of the vortices in the whole superlattice system. Meanwhile, an antivortex state is also shown in the STO layers, perfectly aligned with the vortex cores.

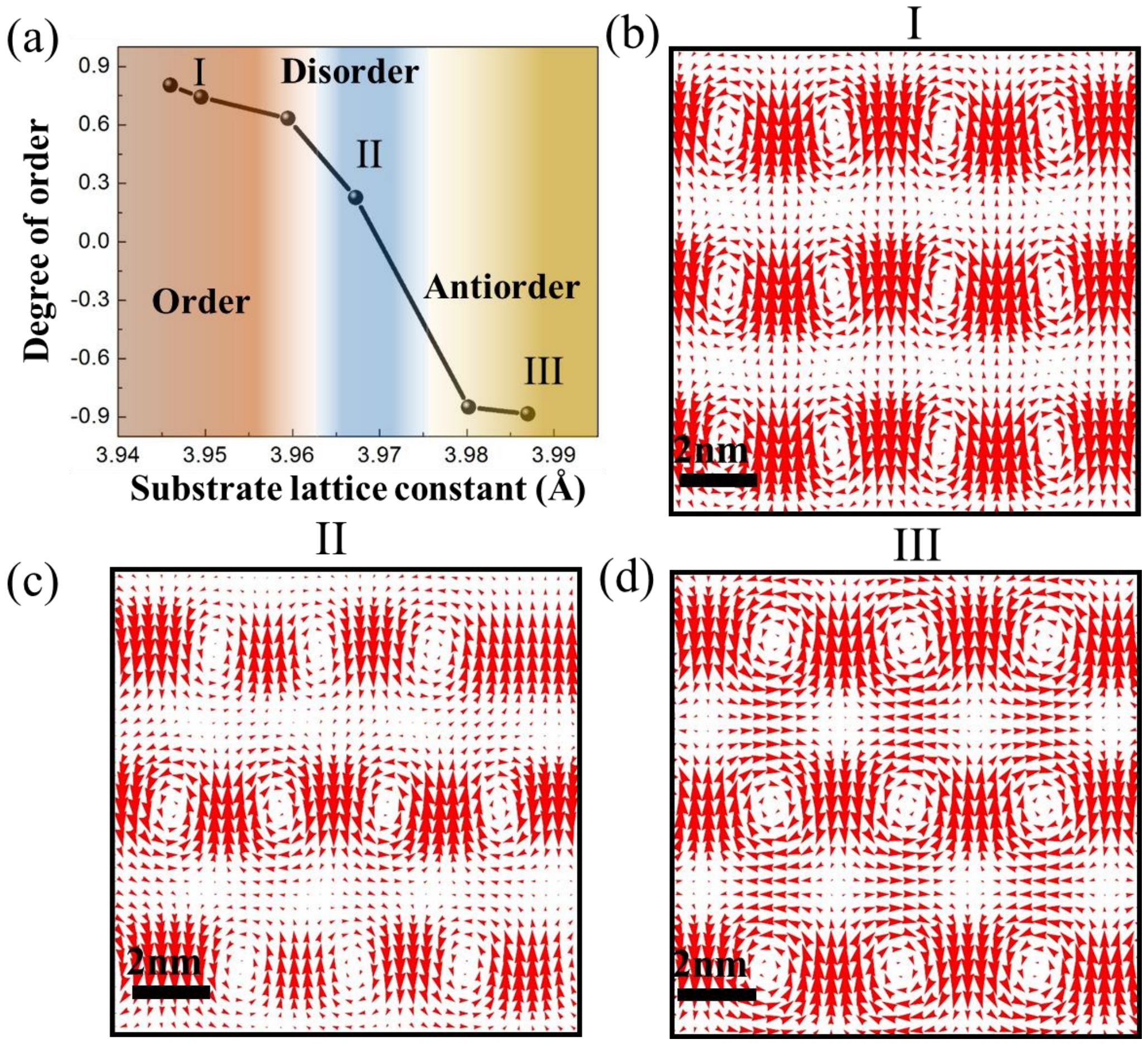

**Fig. 4| Strain phase diagram for $(PZT)_6/(STO)_6$ superlattice**. (a) The degree of order as a function of substrate in-plane lattice constant, indicating the formation of three different phases, i.e., order, disorder and antiorder states. (b) Order state with substrate lattice constant of 3.95 Å. (c) Disorder state with substrate lattice constant of 3.966 Å. (d) Anti-order state with substrate lattice constant of 3.987 Å.

When the in-plane lattice constant is large enough, the compressive strain on the PZT layer is reduced and the epitaxial strain on the STO increases. The in-plane-component of the polarization is the main physical ingredient to determine the ground state of the system. As already discussed in **Fig. 2**, the antiordered structure is the most stable phase, as seen in **Fig. 4(d)**, with the antivortex in the STO layer now located midway between two consecutive vortices.

In the intermediate regime, a compromise between the two previous scenarios is found. It can be shown how the polarization in STO layers tilts, which is neither horizontal nor perpendicular. The tilting of polarization is largely due to the competition of strain effect (which tends to orient the polarization in the in-plane direction with tensile strain) and the poling effect by the PZT layer (which tends to orient the polarization to the out-of-plane direction). As shown in **Fig. 4(c)** and **Fig. S7,** in this case vortices in the neighboring ferroelectric layers can have both order, antiorder and weak correlation regions, which forms a *vortex-glass-like* state with only short-range ordering.

To gain physical insights into the strain-induced order-disorder-antiorder phase transition, the energetics of the three different states with varying substrate lattice constants are investigated. As shown in **Fig. 5**, the relative total energy densities (defined as the total energy density of the polar state minus the total energy density of the centrosymmetric state) of the three structures increase with increasing substrate lattice constant. These three curves intersect with each other, giving rise to three energetically more favorable polar states at different strain states, i.e., the ordered structure has a lower energy density when substrate lattice constant is less than 3.963 Å, while the lower energy state with larger substrate lattice parameter (from 3.975 Å to 4 Å) is the antiordered state, whereas the disordered structure is more favorable in between (3.963 Å to 3.975 Å). The increase in the relative energy density can be understood by further plotting the relative elastic energy, which shows a similar trend with the same slope as the total energy for the three states since the other energies are insensitive to the strain condition. Interestingly, the ordered structure is elastically favorable over a larger strain range (i.e., with substrate lattice constant smaller than 3.99 Å). The other energy densities (e.g., electric, Landau, and gradient energy densities) are given in **Table S1**. It is clearly shown that the sum of energies other than elastic energy densities is higher for the ordered structure, followed by the disorder state. Thus the stable ordered structure is expected only at a lower substrate lattice constant range. The higher electric energy of the ordered structure is due to the formation of electrically-unfavorable *c*-like regions in the STO layers. Whereas the larger gradient energy density of the ordered state is a result of the higher density of vortex arrays.

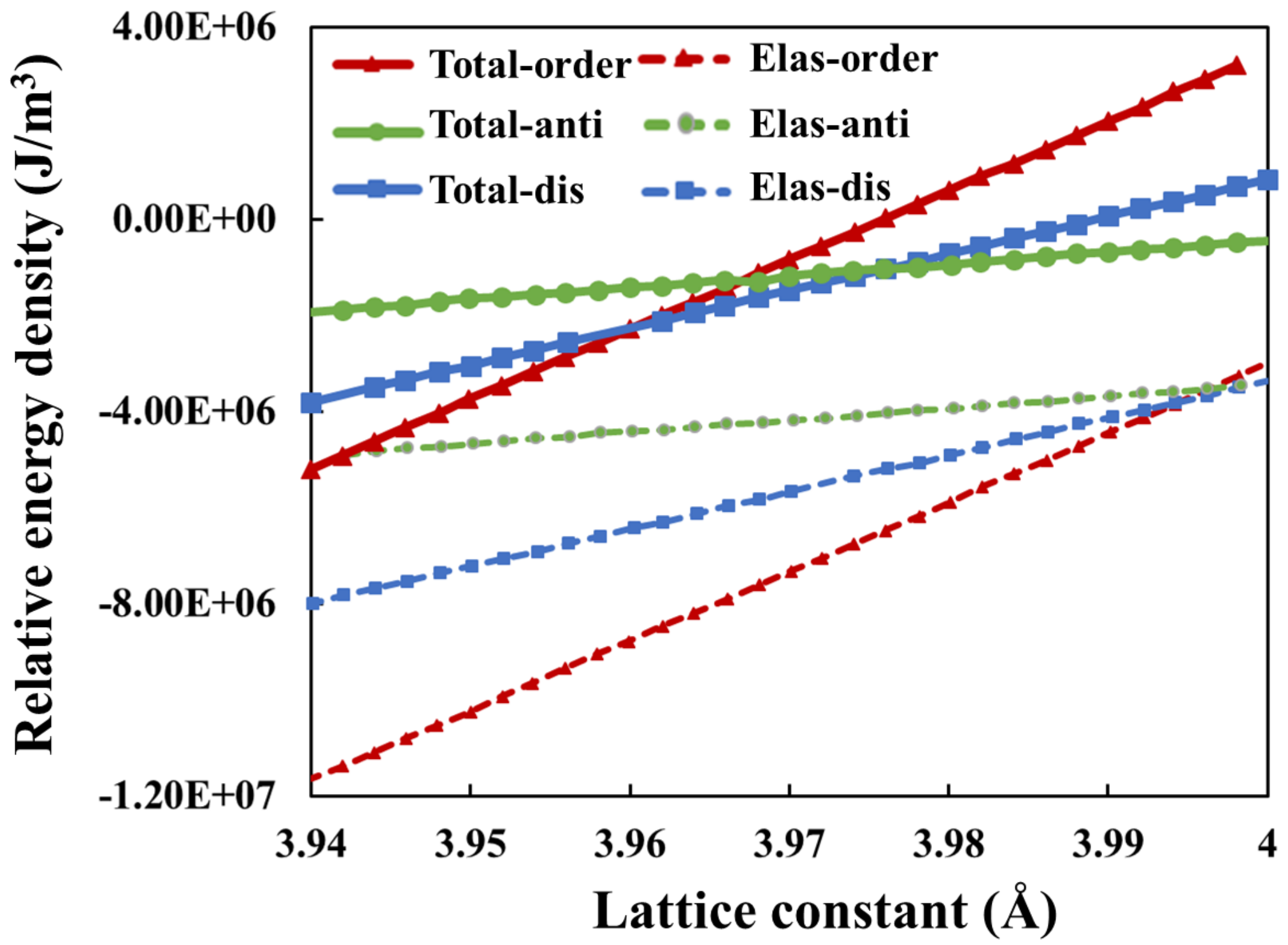

**Fig. 5| Energetics of the strain-mediated phase transitions**. The relative total energy density and elastic energy density with respect to substrate strain.

The relative elastic energy densities of the PZT and STO layers are separated (**Fig. S8**). It is shown that in the PZT layers, all three states show an increase in the relative elastic energy densities with an increase in the substrate lattice parameter. A larger slope is observed for the ordered phase since the polarization in this phase is larger than in other phases. On the other hand, for the STO layers, the ordered phase shows an increase in the relative elastic energy since the *c*-type domain is not elastically favorable for STO layers with largely tensile strain. Meanwhile, the large in-plane polarization could lower the elastic energy of the STO layers, giving rise to a lowering of the elastic energy in these layers for the antiorder and the disorder states. Since the polarization in the STO layers is relatively small, the eigen strains (defined as the phase transition strain due to the electrostriction effect) of the STO layers are also correspondingly smaller, showing a smaller slope in the decreasing trend of the elastic energy in these layers.

In conclusion, we investigated the polar structure in the $(PZT)_n/(STO)_n$ superlattice systems through phase-field simulations. An antiorder state is revealed for short periodicity superlattice on an SSO substrate, where the vortices in the neighboring layers anti-align with each other, resulting from the in-plane polarization in STO layers due to the large tensile strain. An antivortex emerges within STO, located in between two consecutive vortices. Increasing the periodicity leads to the antiorder to disorder transition, owing to the tilt of the domains of the head-to-head and tail-to-tail domains within the STO layer. Keeping the periocitiy of the superlattice constant and sufficiently small, a strain-mediated order-disorder-antiorder transition is revealed, due to the intimate competition between the induction of out-of-plane (due to interfacial coupling), and in-plane (due to strain) in the STO layer. We hope this study could spur further interest towards the understanding of order-disorder transitions for ferroelectric topological structures.

**Acknowledgements**

ZH gratefully acknowledge a start-up grant from Zhejiang University. This work is supported by the Fundamental Research Funds for the Central Universities (No. 2021FZZX003-02-03, ZH). The financial support from Grant PGC2018-096955-B-C41 funded by MCIN/AEI/10.13039/501100011033 is acknowledged (JJ, PGF, FGO). FGO acknowledge financial support from Grant No. FPU18/04661 funded by Spanish Ministry of Universities. The phase-field simulation is performed on the MoFang III cluster on Shanghai Supercomputing Center (SSC).